\documentclass[a4paper]{article}
\usepackage{indentfirst}
\usepackage{graphicx}
\usepackage{setspace}
\title{\mbox{MINOS Anomaly as A Signal of Lorentz Violation}}
\author{Chun Liu$^1$, Jin-tao Tian$^2$, Zhen-hua Zhao$^1$\\$^1$Institute of Theoretical Physics, Chinese Academy of Sciences,\\
and State Key Laboratory of Theoretical Physics,\\
P.O. Box 2735, Beijing 100190, China\\
$^2$National Astronomical Observatories,\\ Chinese Academy of Sciences,\\ Beijing 100012, China}
\begin{document}
\maketitle
\begin{abstract}
Recently, the MINOS collaboration reported an anomaly that the mass-squared difference and mixing angle of $\bar{\nu}_\mu$ $\leftrightarrow$ $\bar{\nu}_\tau$ are both different from that of $\nu_\mu$ $\leftrightarrow$ $\nu_\tau$. In this letter, based on the framework of neutrino oscillations, terms that break the Lorentz symmetry are used as perturbation to explain this anomaly and satisfactory results are got. Remarkably, some surprising conclusions, one of which is that in the high energy limit (hundreds of GeV) neutrino oscillation pattern will be independent of energy, are also arrived.
\end{abstract}
\setlength{\parskip}{0.5\baselineskip}
\section{Introduction}
Observations on solar neutrinos \cite{1} and atmospheric
neutrinos \cite{2} have provided compelling evidences
for neutrino oscillations. Reactor \cite{3} and accelerator
\cite{4} neutrino experiments have further confirmed
the oscillation paradigm. Nowadays, the fact that neutrinos do
oscillate between different flavors has been established. The
original idea of neutrino oscillation was proposed by Pontecorvo
\cite{5} assuming neutrino-antineutrino oscillation in a
pattern similar to that between $K^0$ and $\bar{K}^0$. This idea
was extended to be among different flavors of neutrinos by Maki,
Nakgawa and Sakada \cite{6}. One of the key point (MSW effect) in the
neutrino oscillation paradigm is due to Wolfenstein
\cite{7}, Mikheyev and Smirnov \cite{8} who pointed
out an effect induced by matter when a neutrino passes through
it and interacts with the particles forming the matter. Now it
is well known that massive neutrinos naturally result in neutrino
oscillations among different flavors. It is also noteworthy that
other mechanisms, like Lorentz symmetry violation
\cite{9,10}, can also accommodate neutrino
oscillations. There have been many papers discussing neutrino oscillations using Lorentz violation in the literature \cite{11}.

Neutrino oscillation data fix the neutrino mass-squared
differences and their mixings. According to a global neutrino oscillation data analysis within the three-flavor framework \cite{12}, the best-fit values of oscillating parameters are given as following, $\Delta m^2_{12}=(7.59^{+0.20}_{-0.18})\times 10^{-5} eV^2$, $\Delta m^2_{13}=(2.45^{+0.09}_{-0.09})\times 10^{-3} eV^2$ and $\sin^2(\theta_{12} ) =0.312^{+0.017}_{-0.015}$, $\sin^2(\theta_{23} ) =0.51^{+0.06}_{-0.06}$, $\sin^2(\theta_{13} )=0.010^{+0.009}_{-0.006}$. Within the accuracy of present experiments, the three flavor oscillations can be reduced to two flavor oscillations in two sectors i.e. the `solar' sector and the `atmospheric' sector. In the two flavor analysis, the oscillation probability can be written as,

\begin{equation}
P=\sin^2(2\theta )\sin^2(\frac{\Delta m^2  L}{4E}).
\end{equation}

However, recently the MINOS collaboration reported an anomaly in $\bar{\nu}_\mu$ disappearance experiment: the oscillation parameters are determined to be $\Delta m^2=(3.36^{+0.45}_{-0.40} )\times 10^{-3}$ $eV^2$ and $\sin^2(2\theta ) =0.86 ^{+0.11}_{-0.11}$ \cite{13}. This result is still consistent with that of $\nu_\mu$ disappearance experiment within the 3$\sigma$ level, but if we take the central value seriously it may imply CPT violation, in comparison with the oscillation parameters \ $\Delta m^2=(2.35^{+0.11}_{-0.08})\times 10^{-3}$ $eV^2$ and $\sin^2(2\theta ) =1.00$ ($\sin^2(2\theta) > 0.91$ at 90\% CL) \cite{14} determined in $\nu_\mu$ disappearance experiment. There have been some attempts to solve this anomaly either using CPT violation \cite{15,16} or in terms of non-standard neutrino interaction \cite{17,18,19,20}, for a review about these attempts see \cite{21} and references therein. In the end of section 3, we will do a detailed comparison between these attempts with ours after having presented our model.

                                                   \section{Formalism and Model}
We consider the MINOS anomaly as a signal of Lorentz and
CPT violation. In Refs. \cite{10,22}, the authors point out that observable neutrino oscillations may be a combined result of neutrino masses and Lorentz violation, and results of some neutrino oscillation experiments even can be explained by Lorentz violation without using mass terms. In this letter, we still work in the conventional massive
neutrino paradigm which solves the solar neutrino and atmospheric
neutrino problems. To explain the MINOS anomaly, a Lorentz and
CPT violating term is included as perturbation. We adopt a framework
called Standard Model Extension (SME) \cite{9,22}. It is the general effective theory constructed from SM and allows any coordinate-independent Lorentz violation, which might arise from the Planck scale physics. In the minimal SME lagrangian, all possible renormalizable terms constructed from SM fields which break the Lorentz symmetry are added to the usual SM lagrangian.

In SME, the effective Hamiltonian in the neutrino sector takes the following form,
\begin{equation}
(H)_{ab}=(m^2)_{ab}/(2E)+(a)_{ab}+(cE)_{ab},
\end{equation}
where $(m^2)_{ab}/(2E)$ is the conventional mass squared term, the other two terms are Lorentz violating, the term $a$ is CPT odd and the term $c$ is CPT even. The effective Hamiltonian for anti-neutrinos can be gained by reversing the sign of $a$. Moreover, $a$ does not change with energy while $cE$ is proportional to energy, so that these two terms can complicate the dependence of neutrino oscillations on energies. In the following, we will explain the MINOS anomaly by including the term $a$ in the Hamiltonian. Thus, we would like to discuss the property of the term $a$ in detail before presenting our model.

The term $a$ has something in common with the matter potential induced by the MSW effect: both of them are independent of energy and CPT-odd \cite{22,23}.
However, the term $a$ has differences with the MSW effect in the following two aspects: on the one hand, the MSW effect can appear only when a neutrino passes through matter and is dependent on the density and ingredient of matter \cite{7,8,23}, while the term $a$ is always constant as a vacuum property \cite{22}; on the other hand, the matter potential only appears diagonally in the flavor basis while the term $a$ may have non-vanishing off-diagonal elements. In addition, the MSW effect does not play any role in the oscillations between $\nu_\mu$ and $\nu_\tau$ when considered in the two flavor analysis, because the matter potential induced by normal matter is proportional to the identity matrix in the ($\nu_\mu,\nu_\tau$) basis. In contrast, we can assume that $a_{\mu\mu}$ differs with $a_{\tau\tau}$.

In the following, we assume that the terms $cE$ and $a_{ex}$ in Eq. (2) are absent or negligible for some reasons. Considering the energy range and the baseline distance in the MINOS experiment, the oscillation between $\nu_e$ and $\nu_\mu$ is negligible, so the two flavor analysis is still a good approximation. Hence, the effective Hamiltonian in the ($\nu_\mu,\nu_\tau$) basis can be written as,
\begin{spacing}{1.5}
\begin{equation}
 H=\left(\begin{array}{cc}
       0&\frac{m^2_1}{2E}-a_1\\
       \frac{m^2_1}{2E}-a_1&\frac{m^2_2}{2E}-a_2
\end{array}
\right).
\end{equation}
\end{spacing}
The term on the top left corner has been chosen to be zero because the mixing only has to do with the difference of the two terms on the diagonal. Besides, the minus signs before $a_1$ and $a_2$ are assigned to insure that the values of $a_1$ and $a_2$ are positive in order to explain the MINOS anomaly as we shall see.

For oscillations between $\bar{\nu}_\mu$ and $\bar{\nu}_\tau$, the effective Hamiltonian will be got by reversing the signs before $a_1$ and $a_2$,

\begin{spacing}{1.5}
\begin{equation}
 H=\left(\begin{array}{cc}
       0&\frac{m^2_1}{2E}+a_1\\
       \frac{m^2_1}{2E}+a_1&\frac{m^2_2}{2E}+a_2
\end{array}
\right).
\end{equation}
\end{spacing}

For convenience, we parameterize Eq. (3) and Eq. (4) as follows,
\begin{equation}
 H=\left(\begin{array}{cc}
       0&a\\
       a&b
\end{array}
\right),
\end{equation}
where $a$ represents $\frac{m^2_1}{2E}-a_1$ and $\frac{m^2_1}{2E}+a_1$ in the neutrino sector and anti-neutrino sector respectively, $b$ represents $\frac{m^2_2}{2E}-a_2$ and $\frac{m^2_2}{2E}+a_2$ in the neutrino sector and anti-neutrino sector respectively. In this case, the oscillation probability can be written as,
\begin{equation}
P=\frac{4a^2}{4a^2+b^2}\sin^2(\frac{\sqrt{4a^2+b^2}}{2}L),
\end{equation}
where $\frac{4a^2}{4a^2+b^2}$ and $\frac{\sqrt{4a^2+b^2}}{2}$ play the role of $\sin^2(2\theta )$ and $\frac{\Delta m^2}{4E}$ in Eq. (1) respectively.

We observe that if $\frac{m^2_2}{2E}$ and $a_2$ cancel at $E$ $\sim$ several GeV, the term on the diagonal will be close to zero in Eq. (3) and the oscillations between $\nu_\mu$ and $\nu_\tau$ will be nearly maximal in this energy range. In contrast, $a_2$ has the opposite sign in Eq. (4), so the mixing between $\bar{\nu}_\mu$ and $\bar{\nu}_\tau$ is not maximal any more. Thus, we can understand why $\sin^2(2\theta)=0.86 ^{+0.11}_{-0.11}$ for anti-neutrinos is smaller than that for neutrinos. Furthermore, both values of $a$ and $b$ in the anti-neutrino sector are larger than that in the neutrino sector, resulting in that the energy eigenvalue difference in the anti-neutrino sector is larger than that in the neutrino sector. Therefore, the differences between results observed in the neutrino sector and anti-neutrino sector can be well understood.

                                         \section{Results and Discussions}
In the MINOS experiment, neutrinos whose energy spectrum mainly ranges from 1 to about 10 GeV have a flight of 735 km before being detected. The flux of neutrinos peaks at 3 GeV and has a mean energy of 4 GeV.

We have performed a chi-squared analysis of the data \cite{13} and the best-fit values of the parameters are determined to be as follows,
\begin{center}
\begin{eqnarray}
      a_1&=&3.8\times 10^{-14} eV,\\
      a_2&=&2.2\times 10^{-13} eV,\\
     m^2_1&=&1.4\times 10^{-3} eV^2,\\
     m^2_2&=&7.9\times 10^{-4} eV^2.
\end{eqnarray}
\end{center}
Using these values, we compare the expectation of events with the data in $\bar{\nu}_\mu$ and $\nu_\mu$ disappearance experiments in Fig. 1 and Fig. 2, respectively.

\vskip 1 cm
 \begin{center}
\includegraphics[scale=0.3]{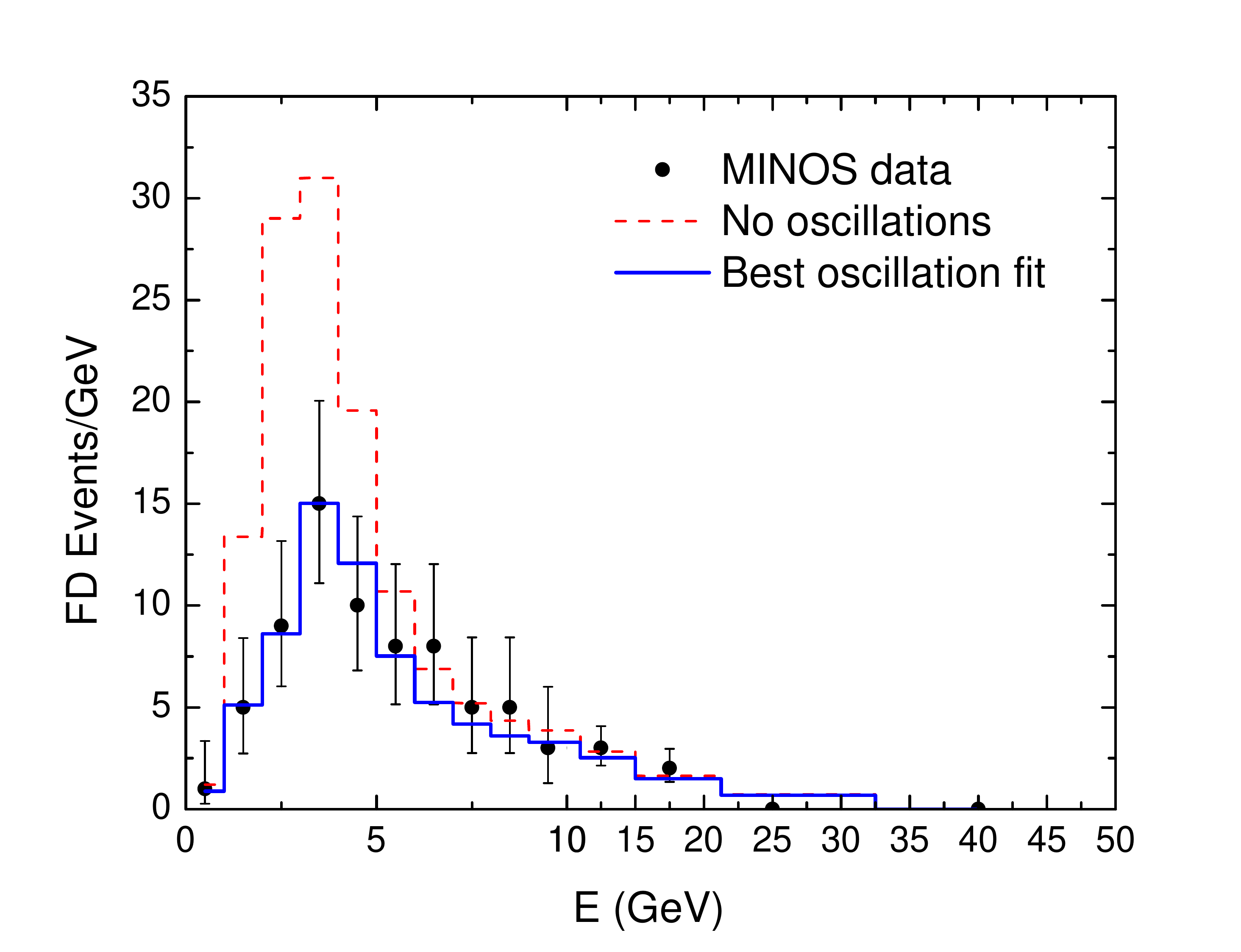}
\end{center}
Fig.1: Comparison of the measured Far Detector $\bar{\nu}_\mu$ energy spectrum to the expectation in two cases: in the absence of oscillation; using the oscillation parameters given in Eqs. (7-10).\\
 \begin{center}
\includegraphics[scale=0.3]{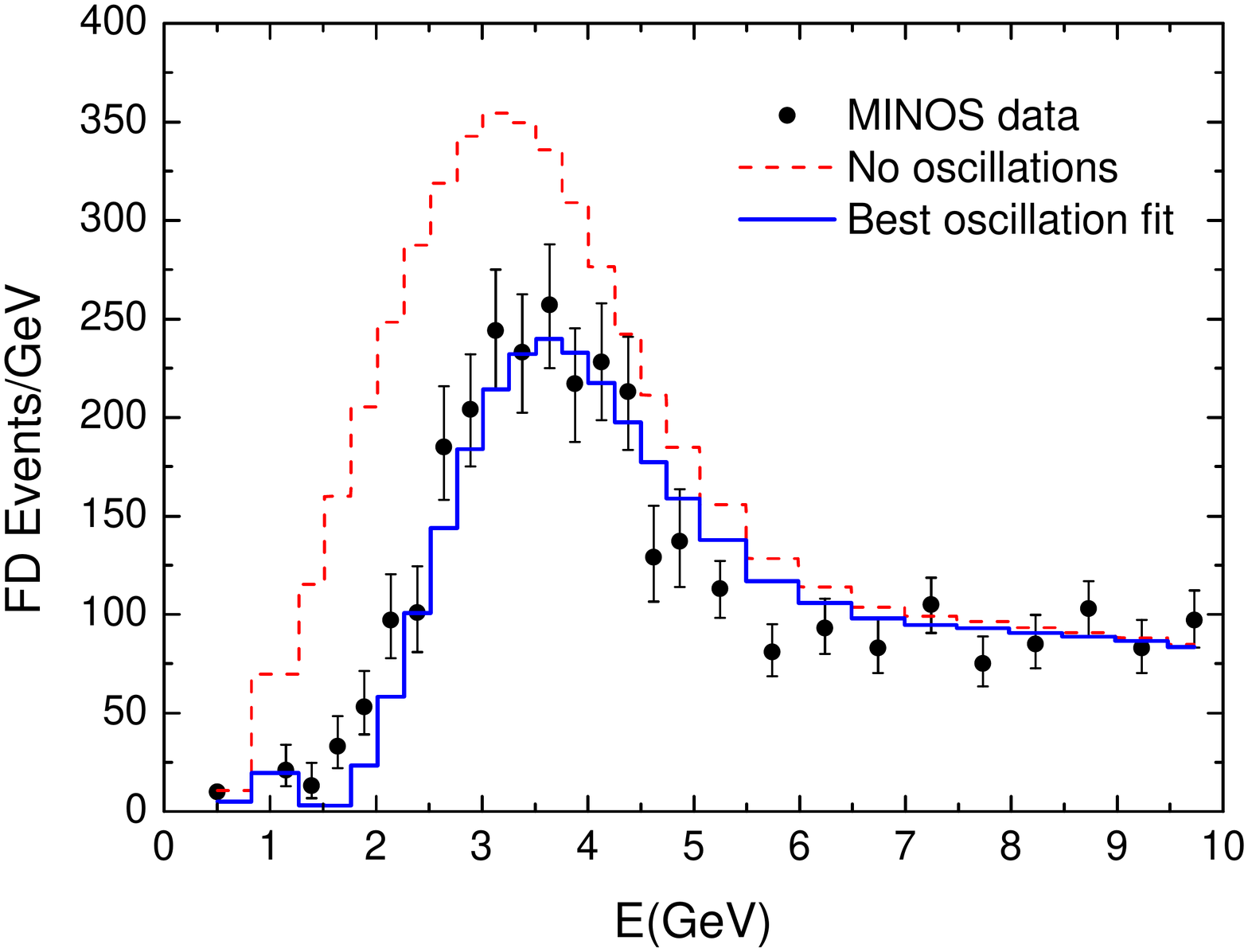}
\end{center}
Fig.2: Comparison of the measured Far Detector ${\nu}_\mu$ energy spectrum to the expectation in two cases: in the absence of oscillation; using the oscillation parameters given in Eqs. (7-10).
\vskip 0.15 cm

In the energy range of the MINOS experiment, it is difficult to distinguish the model in the present letter with the models where the mass matrices in the neutrino sector and anti-neutrino sector are different. However, as energy rises, the mass term contribution will decrease while $a_1$ and $a_2$ do not change, therefore, the two scenarios will show different properties in the high energy limit. When the energy of neutrinos reaches several hundred GeV, the mixing induced completely by mass terms will become very small as shown in Eq. (1) given a fixed distance $L$, while the mixing induced by mass terms plus the Lorentz violation effect will become independent of energy as shown in Fig. 3 and Fig. 4 below.

 \begin{center}
\includegraphics[scale=0.7]{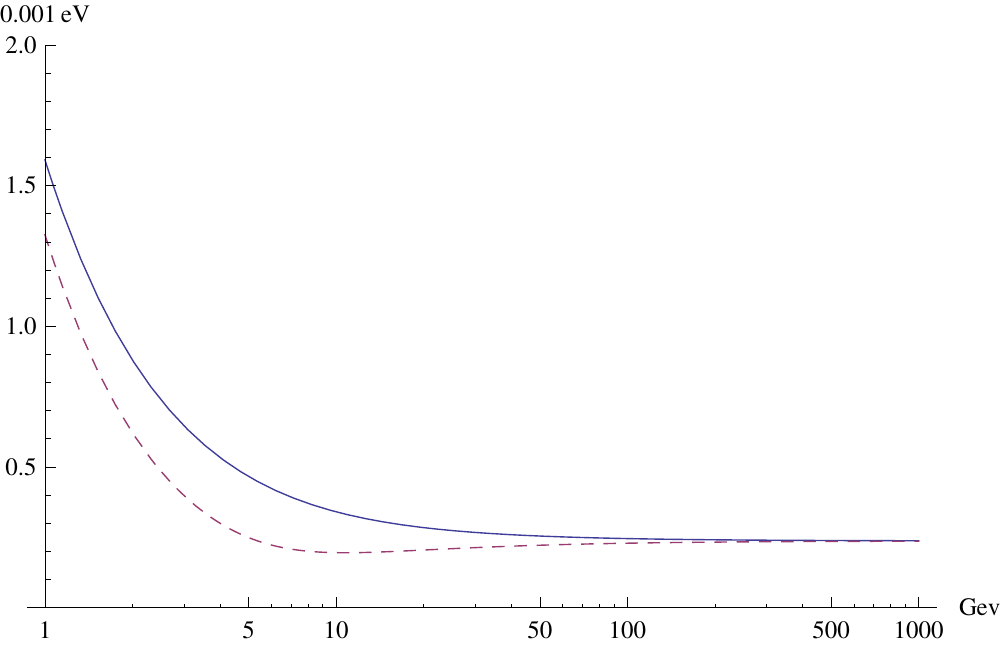}
\end{center}
Fig.3: $\Delta E$ due to Eq. (3) (dashed line) and Eq. (4) (solid line) as a function of energy.
 \begin{center}
\includegraphics[scale=0.7]{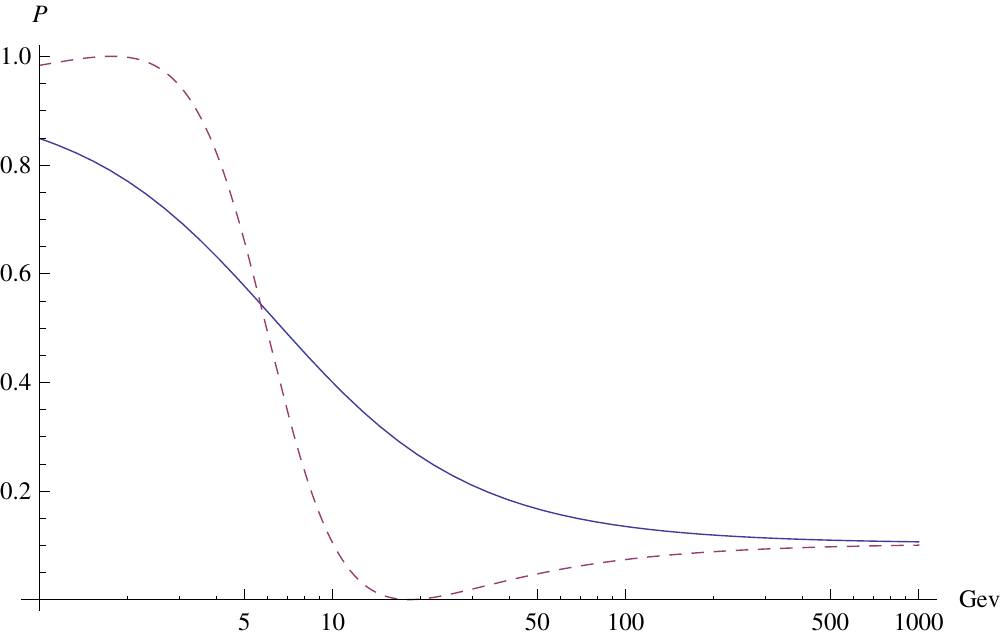}
\end{center}
Fig.4: $\sin^2(2\theta )$ due to Eq. (3) (dashed line) and Eq. (4) (solid line) as a function of energy.
\vskip 0.15 cm
Fig. 3 shows $\Delta E$ as a function of energy in the neutrino sector and anti-neutrino sector; Fig. 4 shows $\sin^2(2\theta )$ as a function of energy in the neutrino sector and anti-neutrino sector. As shown in these two figures, in the high energy limit where $\frac{m^2}{2E} \ll a_1$ and $a_2$, there is almost no difference between $\nu_\mu \leftrightarrow \nu_\tau$ and $\bar{\nu}_\mu \leftrightarrow \bar{\nu}_\tau$ and these two oscillation patterns will be independent of energy. There is a distinct character in the neutrino sector: when the energy of neutrinos is around 20 GeV, $\sin^2(2\theta )$ has a vanishing point, so the oscillations between $\nu_\mu$ and $\nu_\tau$ are highly suppressed in this small energy range.

It should be mentioned that in the energy range of solar neutrinos (a few MeV), $\frac{m^2}{2E}$ $\gg$ $a_1$ and $a_2$, so $a_1$ and $a_2$ will not affect the explanation of the solar neutrino problem. However, this model will be constrained by the atmospheric neutrino oscillation data. In Super-K experiment, events generated by neutrinos and anti-neutrinos can not be distinguished, so the results can not supply strong constraints. In future, the long baseline experiments for the oscillations $\nu_\mu$ $\leftrightarrow$ $\nu_\tau$ and $\bar{\nu}_\mu$ $\leftrightarrow$ $\bar{\nu}_\tau$ at high energy (several tens of GeV) with high precision can confirm or exclude this model.

Finally, we would like to do a comparison between other attempts solving the MINOS anomaly with ours. In \cite{15,16}, the authors use CPT violation to solve the MINOS anomaly and consider that neutrinos and anti-neutrinos do have unequal mass differences at least in the atmospheric neutrino sector. However, in our model, mass differences in the neutrino sector and that in the anti-neutrino sector are equal. Because the mass term and the term $a$ have different dependence on energy, the models in \cite{15,16} which only contain the mass term will show different properties on energy from our model which includes the term $a$. Consequently, experiments in different energy ranges can distinguish these two kinds of models. In the low energy range where the mass term is much larger than the term $a$ in our model, there is no CPT violation observable in neutrino oscillation experiments, while there is always CPT violation irrespective of the energy of neutrinos in the models of \cite{15,16}. Only when the energy of neutrinos reaches the order of GeV can the CPT-violating effect emerge in our model.

When it comes to \cite{17,18,19,20}, we have given the comparison between the term $a$ and the potential induced by neutrino interaction in section 2. They share some similarities, but the essential difference between them is that the CPT-violating term $a$ is property of vacuum, while the matter potential is dependent on the concrete circumstance the neutrinos experience during their flight. Experiments can distinguish these two mechanisms too.

                                                                       \section{Summary}
 In conclusion, the MINOS anomaly if it really exists can be successfully explained by using a CPT-violating term in the formalism of SME as perturbation to the conventional mass-induced oscillation paradigm. Meanwhile, some odd conclusions are arrived: in the energy range around 20 GeV, oscillations between $\nu_\mu$ and $\nu_\tau$ will be highly suppressed due to the very small mixing; in the high energy limit where the mass term is small enough compared to the terms from SME, $\nu_\mu$ $\leftrightarrow$ $\nu_\tau$ and $\bar{\nu}_\mu$ $\leftrightarrow$ $\bar{\nu}_\tau$ will have the same oscillating pattern which is independent of energy. Considering the data are statistically limited, the quantitative results are inconclusive. However, if the MINOS anomaly persists, it maybe become the first evidence of Lorentz violation which has a significant meaning.

                                                                        \section{Acknowledgements}
 One of the authors (Zhao) thanks Xue Chang, Mingkai Du and Jiashu Lu for helpful discussions and other helps in the process of this work. We express our appreciation to Lei Wu for his helps in data analysis. This work was supported in part by the National Natural
Science Foundation of China under Nos. 90503002, 10821504 and 11075193 and by
the National Basic Research Program of China under Grant No. 2010CB833000.

\end{document}